# Calorimetric evidence for enhancement of homogeneity in high performance $Sr_{1-x}K_xFe_2As_2$ superconductors

Chiheng Dong[a], Chao Yao[a], He Huang[a], Xianping Zhang[a], Dongliang Wang[a], Yanwei Ma[a,b,*]

[a]Key Laboratory of Applied Superconductivity, Institute of Electrical Engineering, Chinese Academy of Sciences, Beijing 100190
[b]University of Chinese Academy of Sciences, Beijing 100049, People's Republic of China
*Corresponding author: Tel.: +86-10-82547129; e-mail: ywma@mail.iee.ac.cn

**Abstract:**
We comparatively studied the critical current density, magnetization and specific heat of the rolled and the hot-pressed $Sr_{1-x}K_xFe_2As_2$ tapes. The Schottky anomaly that is obvious in the specific heat of the rolled tape disappears in the hot-pressed tape. Moreover, the hot-pressed tape has a higher fraction of superconductivity and a narrower distribution of superconducting transition temperature than the rolled tape. Combined with the magnetization data, we conclude that sintering under high pressure provides a better environment for complete chemical reaction and more homogenous dopant distribution, which is beneficial to the global current of a superconductor.

The newly discovered iron-based superconductors [1] provide not only a new platform to investigate the superconducting mechanism but also an opportunity to propel the practical application. Their parent compounds usually undergo an antiferromagnetic (AFM) transition accompanied by a structural transition. Isovalent or aliovalent substitutions suppress the AFM order, and superconductivity prevails thereafter. It naturally raises a question on whether the dopants are homogeneously distributed in the iron-based compounds. There are many papers reporting that there is an electronic or chemical phase separation in the iron-based superconductors [2-5]. For example, the Se and Te atoms in the optimally doped $FeTe_{1-x}Se_x$ superconductors are not uniformly distributed but aggregate into Te and Se patches [6,7]. This chemical phase separation greatly affects the macroscopic properties at normal and superconducting state [8]. Similarly, the K atoms in $(Ba/Sr)_{1-x}K_xFe_2As_2$ cluster into K-rich and K-poor regions, resulting in a nano-scale variation of superconducting gap [9,10]. Due to sensitive dependence on $x$, the superconducting transition temperatures $T_c$ are also different in these regions. Moreover, the critical current densities $J_c$ of these regions depend on the doping level because upper critical field $H_{c2}$, correlation length $\xi$ and flux pinning landscape change greatly with $x$ [11]. As a result, this chemical phase separation will affect the overall $J_c$ of a superconductor.

For large-scale application, the iron-based superconductors must be made into wires and tapes via the powder in tube method [12]. Recently, the $J_c$ of the hot pressed (HP) $Sr_{1-x}K_xFe_2As_2$ tapes at 4.2 K and 10 T has surpassed the practical application level of $10^5$ A/cm$^2$ [13]. The enhancement of $J_c$ is attributed to the high core density, good texture [14] and strong flux pinning [15]. The energy dispersive spectroscopy mapping indicates a uniform elements distribution [16]. However, due to the limited accuracy and spatial resolution [17], some microscopic difference may be overlooked. Specific heat measurement reflects bulk properties of a superconductor that gives important information about quasiparticle density of states, superconducting gap and $T_c$ distribution. Here we measure the specific heat of the rolled and the HP $Sr_{0.6}K_{0.4}Fe_2As_2$ superconducting tapes to study the phase homogeneity and its relation to $J_c$. The precursor was prepared by a solid state reaction method according to the nominal composition $Sr_{0.6}K_{0.4}Fe_2As_2$. The precursor was then packed into a silver tube, swaged, drawn and flat rolled into tapes. The rolled tape was hot-pressed into the HP tape at 880 ℃ under 30 Mpa. The transport critical current was measured by a four-probe method. We removed the silver sheath and measured the susceptibility and the specific heat of the superconducting core on a physical properties measurement system.

Fig. 1 shows the magnetic field dependence of transport critical current density. Both the rolled and the HP tapes show a robust field dependence, while the $J_c$ of the HP tape is about three times larger than that of the rolled tape. The inset shows the temperature dependence of normalized susceptibility. The superconducting transition near $T_c$ of the HP tape is sharper than that of the rolled tape, manifesting that the superconducting phase of the HP tape is more homogeneous.

Fig. 2(a) and (b) show the low temperature specific heat C/T as a function of $T^2$ at zero field. Generally, the specific heat of a superconductor at a low temperature can be expressed as $C=\gamma_0 T +C_{phonon}+C_e+C_{Sch}$, where $\gamma_0 T$ represents the residual electronic specific heat that comes from the non-superconducting fraction of the sample [18] and $C_{phonon}=\beta_3 T^3+\beta_5 T^5+\beta_7 T^7+\ldots$ is the phonon term. The superconducting electron contribution $C_e$ is given by $C_e=Dexp(-\Delta(0\ K)/k_B T)/T^{1.5}$ based on an $s$-wave paring symmetry [19], where $k_B$ is the Boltzmann constant, $\Delta(0\ K)$ is the superconducting gap at 0 K. The Schottky anomaly $C_{Sch}$ generally results from lifting the degeneracy of the states of the paramagnetic spins from the magnetic impurities [19]. In a two-level system ($S$=1/2) [20], the Schottky anomaly $C_{sch}$ can be simplified into:

$$C_{Sch}=n(g\mu_B H_{eff}/k_B T)^2 \frac{exp(g\mu_B H_{eff}/k_B T)}{(1+exp(g\mu_B H_{eff}/k_B T))^2}\ , \qquad (1)$$

where $n$ is the concentration of the paramagnetic centers, $g$ is the Landé factor, $\mu_B$ is the Bohr magneton and $H_{eff}$ is the effective magnetic field.

For the rolled tape, the $C/T$-$T^2$ curve shows a linear behavior down to 3.7 K and an upturn afterwards. This upturn is the so-called Schottky anomaly. We firstly deal with the data between 3.7 K and 7 K where $C_e$, $C_{sch}$, the quintic term and the higher power term of the phonon contribution can be neglected. We thus fit the curve using $C/T=\gamma_0+\beta_3T^2$, as shown by the black dashed line in Fig. 2(a). The fitted parameters are $\gamma_0=7.35$ mJ/mol K$^2$, $\beta_3=0.63$ mJ/mol K$^4$. We removed the phonon term $C_{phonon}/T \sim \beta_3T^2$ from $C/T$, and observe an obvious upturn at a low temperature as shown in the inset of Fig.2 (a). We fit the $(C-C_{phonon})/T$-$T^2$ curve up to 14 K using the equation below:

$$(C-C_{phonon})/T=\gamma_0+C_e/T+C_{sch}/T. \qquad (2)$$

We fix $\gamma_0=7.35$ mJ/mol K$^2$ and set other parameters free. The obtained superconducting gap at 0 K is $\Delta(0\ K)=4.72$ meV, D=$3.2\times10^5$ mJ K$^{0.5}$/mol. Using the parameters obtained above, we fit the $C/T$-$T^2$ curve up to 12 K by using the function:

$$C/T=\gamma_0+C_e/T+\beta_3T^2+\beta_5T^4+C_{Sch}/T, \qquad (3)$$

where $\beta_5=6.97\times10^{-5}$ mJ/mol K$^6$.

For the HP tape, the $C/T$ at a low temperature exhibits a linear dependence on $T^2$ down to 2 K. We thus ignore the $C_{Sch}$ term and fit the curve up to 12 K by using $C/T=\gamma_0+\beta_3T^2+\beta_5T^4+C_e/T$, as shown by the red line in Fig. 2(b). The fitting parameters are $\gamma_0=3.53$ mJ/mol K$^2$, $\beta_3=0.64$ mJ/mol K$^4$, $\beta_5=5.35\times10^{-5}$ mJ/mol K$^6$, D=$6\times10^5$ mJ K$^{0.5}$/mol. The Debye temperature $\Theta_D$ can be derived from the relation $\beta_3=2.4\pi^4 k_B N_A Z\Theta_D^{-3}$, where $N_A$ is the Avogadro constant and Z is the number of atoms per unit cell. The $\Theta_D$ is determined to be 247 K, the superconducting gap at 0 K is $\Delta(0\ K)=5.32$ meV. These two parameters are close to that of the Ba$_{0.6}$K$_{0.4}$Fe$_2$As$_2$ single crystals [19].

Fig. 2(c) and (d) present the specific heat up to 45 K. A pronounced $\lambda$ shape peak of $C/T$ indicates bulk superconductivity. After applying a magnetic field of 9 T, the anomalies are suppressed down to a lower temperature and become less obvious. In order to extract the electronic contribution, the phonon term must be removed. However, it is quite challenging to obtain the phonon term because our highest achievable magnetic field is not sufficient to suppress superconductivity. As a result, we try to fit the normal-state specific heat above T$_c$ at zero field using the following polynomial function [19]:

$$C_{fit}/T= \gamma_n+\beta_3T^2+\beta_5T^4+ \beta_7T^6+ \beta_9T^8+ \beta_{11}T^{10} \qquad (4)$$

by imposing the conservation of entropy, i.e.

$$\int_0^{T_c^{max}} \frac{C_{fit}(T)}{T}dT=\int_0^{T_c^{max}} \frac{C(T)}{T}dT, \quad (5)$$

where $\gamma_n=\gamma_0+\gamma_{ne}$, $\gamma_{ne}$ is the coefficient of the normal state electronic specific heat.

During the fitting, we fixed the $\beta_3$ and $\beta_5$ to the values obtained above and set other parameters free. The fitting lines for the data at 0 T are shown in Fig. 2(c) and (d) in red. The fitting parameters of the rolled tape at 0 T are $\gamma_n$= 43.44 mJ/mol K$^2$, $\beta_7$= -4.93×10$^{-7}$ mJ/mol K$^8$, $\beta_9$= 2.52×10$^{-10}$ mJ/mol K$^{10}$, $\beta_{11}$= -4.20×10$^{-14}$ mJ/mol K$^{12}$. For the HP tape, $\gamma_n$= 37.65 mJ/mol K$^2$, $\beta_7$= -3.49×10$^{-7}$ mJ/mol K$^8$, $\beta_9$= 1.84×10$^{-10}$ mJ/mol K$^{10}$, $\beta_{11}$= -3.08×10$^{-14}$ mJ/mol K$^{12}$. The insets in (c) and (d) are the entropy difference between the normal and the superconducting state. The $S_n$-$S_s$ value drops to zero when temperature increases to above $T_c$, proving that the fitting is reliable. One can see that the normal state C/T-T curves at 0 and 9 T nearly overlap with each other, manifesting that the phonon term does not change much under magnetic field. So it is reasonable to fit the curves at different fields with the same $\beta_3$ and $\beta_5$ as that at zero field.

Fig. 2(e) and (f) show the electronic contribution of specific heat $C_e$/T as a function of temperature at zero field. The Schottky-type contribution has been subtracted from the data of the rolled tape. The entropy conservation law is naturally satisfied as shown in the insets of (e) and (f). The upper black dashed line in the main panel represents $\gamma_n=\gamma_0+\gamma_{ne}$. As the temperature decreases, the specific heat starts to ascend at the same temperature, but the jump seems sharper for the HP tape. The specific heat anomaly $\Delta C/T|_{T_c}$ is estimated to be 67.1 mJ/mol K$^2$ for the rolled tape, lower than the HP sample's value $\Delta C/T|_{T_c}$~71.33 mJ/mol K$^2$. When the temperature decreases further down to 2 K, $C_e$/T does not extrapolate to zero at 0 K but to a residual value $\gamma_0$. The ratio of $\gamma_0$ to its normal-state counterpart $\gamma_n$ yields an estimate of fraction of superconductivity FS=1-$\gamma_0/\gamma_n$ [18]. The actual FS value for the rolled and the HP tapes is 83% and 91% respectively, indicating that the superconducting fraction increases after hot pressing.

A. Junod et al. firstly proposed a method to analyze the Tc of the A15-type compounds [21]. C. Senatore et al. then applied this method to analyze the Tc distribution of the Nb$_3$Sn wires [22-24] and MgB$_2$ bulk samples [25]. This method is also proved to be applicable to the SmFeAsO$_{0.85}$F$_{0.15}$ polycrystals [26] and the (Ba$_{1-x}$K$_x$)Fe$_2$As$_2$ single crystals [9]. Provided that there is a distribution of $T_c$ from 0 K to $T_c^{max}$ with a weight function $f(T_c)$, the electronic specific heat and entropy at a certain temperature can be expressed as [24]:

$$C_e(T)=\int_T^{T_c^{max}} f(T_c)C_{es}(T,T_c)dT_c + \int_0^T f(T_c)C_{en}(T,T_c)dT_c \quad (6),$$

$$S_e(T)=\int_T^{T_c^{max}} f(T_c)S_{es}(T,T_c)dT_c + \int_0^T f(T_c)S_{en}(T,T_c)dT_c \quad (7).$$

The first part on the right side of the equations are the specific heat (6) and the entropy (7) of superconducting electrons, and the second part is the specific heat (6) and the entropy (7) of the electrons at the normal state. We make two assumptions:

(1) all the parts with different $T_c$ have the same electronic specific heat coefficient at normal state: $C_{en}(T,T_c)=\gamma T$;

(2) electronic specific heat in the superconducting state can be described by a generalized two-fluid model: $C_{es}(T, T_c)=n\gamma T_c(T/T_c)^n$.

We can finally obtain the following equation by substituting the above expression of $C_{en}(T, T_c)$ and $C_{es}(T, T_c)$ into Equ. (6) and (7):

$$F(T) \equiv \int_0^T f(T_c)\, dT_c = \frac{nS_e(T)-C_e(T)}{(n-1)\gamma T}. \qquad (8)$$

$F(T)$ is the fraction of the sample with $T_c \leq T$ and $n$ is equal to 3 in the Gorter-Casimir model. $F(T)$ is equal to unity when $T_c \geq T_c^{max}$.

The inset of Fig. 3(a) shows the temperature dependence of $F(T)$ curves at 0 T. The $F(T)$ curves stay at zero below $T_c^{min}$, then undergo a sudden jump and remain to be unity when $T \geq T_c^{max}$. The transition of the HP tape is sharper and starts at a temperature higher than the rolled tape. The main panel of Fig. 3 shows the $f(T)$ curves obtained by $f(T)=dF(T)/dT$ at 0-9 T. We define the maximum and minimum $T_c$ of the tape as the temperature at which $f(T)=0$. We find that the rolled and the HP tapes share the same $T_c^{max}\sim 37$ K, which is consistent with the reported value of the optimally doped $Sr_{1-x}K_xFe_2As_2$ [27,28]. Similar to $NbSn_3$ [22,24,29] and $MgB_2$ [30] tapes, the $Sr_{1-x}K_xFe_2As_2$ tape have a wide $T_c$ distribution rather than a unified $T_c$. The $T_c$ distribution of the HP tape concentrates at 35.5 K and 93 % of it has $T_c$ above 34 K. On the contrary, only 20 % of the rolled tape has $T_c$ above 34 K. In order to quantitatively analyze the $T_c$ distribution, we fit the curves using the multiple Gaussian functions. The $f(T)$ curves are contributed by two sets of Gaussian functions. The one at higher temperature is denoted as peak 1 (H1 for the HP tape and R1 for the rolled tape), the other at lower temperature is denoted as peak 2 (H2 for the HP tape and R2 for the rolled tape). At zero field, the intensities and the positions of H1 and H2 are close to each other. While H2 gradually weakens and shifts away from H1 with the increasing field, the $f(T)$ curve is mainly contributed by H1. The situation is just the opposite for the rolled tape, the $f(T)$ curve is dominated by R2. R1 even disappears at 9 T.

The field dependence of the positions of the Gaussian peaks are summarized in Fig. 4 (a). We can see that the peak positions decrease monotonically with the field. Moreover, the peak positions of H1 and R1 are close to each other. Although there is a 1.5 K distance between the peak positions of H2 and R2 at zero field, this gap is gradually suppressed with the increasing field. Fig. 4(b) shows field dependence of the area ratio of the Gaussian peaks. The area ratio of R1 and R2 is about (20%:80%). On the contrary, H1 occupies nearly 80 % of the HP tape above 0 T. As a result, we only compare the full-width-half-maximum (FWHM) of the Gaussian peaks which make the most contribution to the

*f*(T) curves. As shown in Fig. 4(c), the FWHM of H1 barely changes with the field, remaining between 1 and 1.6. The FWHM of R2 is always larger than that of H1 at 0-9 T, indicating that the $T_c$ distribution of the HP tape is more homogeneous than the rolled tape. It is further corroborated by the $T_c$ distribution width $\Delta T_c = T_c^{max} - T_c^{min}$ as shown in Fig. 4(d). At 0 T, $\Delta T_c$ is 5.9 K and 3.7 K for the rolled and the HP tapes respectively. The $\Delta T_c$ of the HP tape quickly increases to about 6 K at 3 T and saturates afterwards, while the $\Delta T_c$ of the rolled tape increases linearly with the field and is always larger than that of the HP tape.

The ideal condition of a superconductor is that every part of it has a uniform doping level with the highest $J_c$. However, the dopants are intrinsically not homogeneous in $Sr_{1-x}K_xFe_2As_2$ according to our results. The nonuniformly distributed K atoms segregate the superconductor into regions with different doping levels. These regions have different $J_c$ with distinct field dependence, causing $J_c$ fluctuations along the tape. It is not conducive to the overall performance because the critical current $I_c$ is limited by the region with the lowest $J_c$. Fig. 4(e) is a schematic showing qualitatively the doping dependence of $T_c$ and $J_c$. The peak position of $J_c(x)$ locates in the slightly underdoped region [11,31,32]. Both the rolled and the HP tapes are fabricated using the same precursor with nominal composition of $Sr_{0.6}K_{0.4}Fe_2As_2$, so they both attain the highest $T_c$. However, the rolled tape has a wider $T_c$ distribution (shown as the light red bar) than the HP tape (shown as the light blue bar). The narrow $T_c$ distribution corresponds to the larger $J_c^{min}$, and accordingly larger overall transport current than the wide $T_c$ distribution. Therefore, we suggest that narrow $T_c$ distribution and homogeneous superconducting phase are beneficial to global current. Recently, we prepared the $Ba_{1-x}K_xFe_2As_2$ precursor via a two-steps method to make the distribution of Ba and K more homogeneous [33]. A remarkable enhancement of $J_c$ is achieved. Here, we utilize the hot pressing method to ameliorate the dopant clustering in $Sr_{1-x}K_xFe_2As_2$ and obtain a narrower $T_c$ distribution. Similar result was found in the $MgB_2$ compounds. High temperature, high-pressure synthesis is also helpful to create homogeneously C-substituted $MgB_2$ [34]. We thus believe that sintering under high pressure is an effective way to make dopants more homogeneous than traditional synthesis methods. This is important because inhomogeneity may obscure the intrinsic properties and limit the practical application.

In conclusion, we perform a comparative study of heat capacity on the rolled and the HP $Sr_{1-x}K_xFe_2As_2$ tapes. The Schottky anomaly originating from the magnetic impurities in the rolled tape, leads to a smaller fraction of superconductivity than the HP tape. The *f*(T) curves derived via a deconvolution method have an asymmetric peak shape which can be well fitted by the multiple Gaussian functions. We find that most part of the HP tape has a $T_c$ higher than the rolled tape. Moreover, the rolled tape has a wider $T_c$ distribution than the HP tape at 0-9 T. Combined with the magnetization data, we suggest that sintering under high pressure leads to a more complete chemical reaction and a more homogeneous superconducting phase, which is beneficial to the global current in superconductors.

This work is partially supported by the National Natural Science Foundation of China (Grant Nos. 51402292 and 51320105015), the Beijing Municipal Science and Technology Commission (Grant No. Z171100002017006), Bureau of Frontier Sciences and Education, Chinese Academy of Sciences (QYZDJ-SSW-JSC026).

Captions:

Figure 1
Transport critical current densities as a function of magnetic field. The inset shows the temperature dependence of normalized susceptibility with field cooling and zero-field cooling process.

Figure 2
Temperature dependence of specific heat plotted as $C/T$-$T^2$ for (a) the rolled tape and (b) the HP tape at zero field. The inset in (a) shows the specific heat $(C-C_{phonon})/T$ as a function of $T^2$, the blue curve is the fitting line using equ. (2). (c) and (d) is the specific heat $C/T$-$\gamma_0$ as a function of temperature up to 45 K. The insets show the entropy difference between the normal and the superconducting state. (e) and (f) show the electron specific heat as a function of temperature. The upper black dashed line represents $\gamma_n=\gamma_0+\gamma_{ne}$, the lower dashed line represents the residual electron contribution $\gamma_0$. The inset shows the normal and the superconducting state electronic entropies.

Figure 3
$T_c$ distribution functions $f(T)$ for the Rolled (black circle) and the HP (blue square) tapes at 0-9 T. The F(T) curves at 0 T are shown in the inset of (a). The $f(T)$ cures can be well fitted by the multiple Gaussian peaks function. The green and red lines are the Gaussian peaks functions for the HP tape and rolled tape, respectively. The black and blue lines are the envelope sum of the two Gaussian peaks. The Gaussian peaks of the HP(rolled) tape at high and low temperatures are denoted as H1(R1) and H2(R2), respectively. The arrows mark $T_c^{max}$ and $T_c^{min}$, which are defined as the temperatures at which $f(T)=0$.

Figure 4
Field dependence of (a) positions, (b) area ratio and (c) FWHM of the Gaussian peaks. (d) is the $T_c$ distribution width $\Delta T_c$ as a function of magnetic field. (e) is a schematic explaining why the narrower $T_c$ distribution is beneficial to $J_c$. The black and green lines are the $x$ dependent of $T_c$ and $J_c$ respectively. The light blue and light red areas correspond to the narrow and wide $T_c$ distributions, respectively. The corresponding minimum critical current density $J_c^{min}$ is marked by the dashed line.

Figure 1

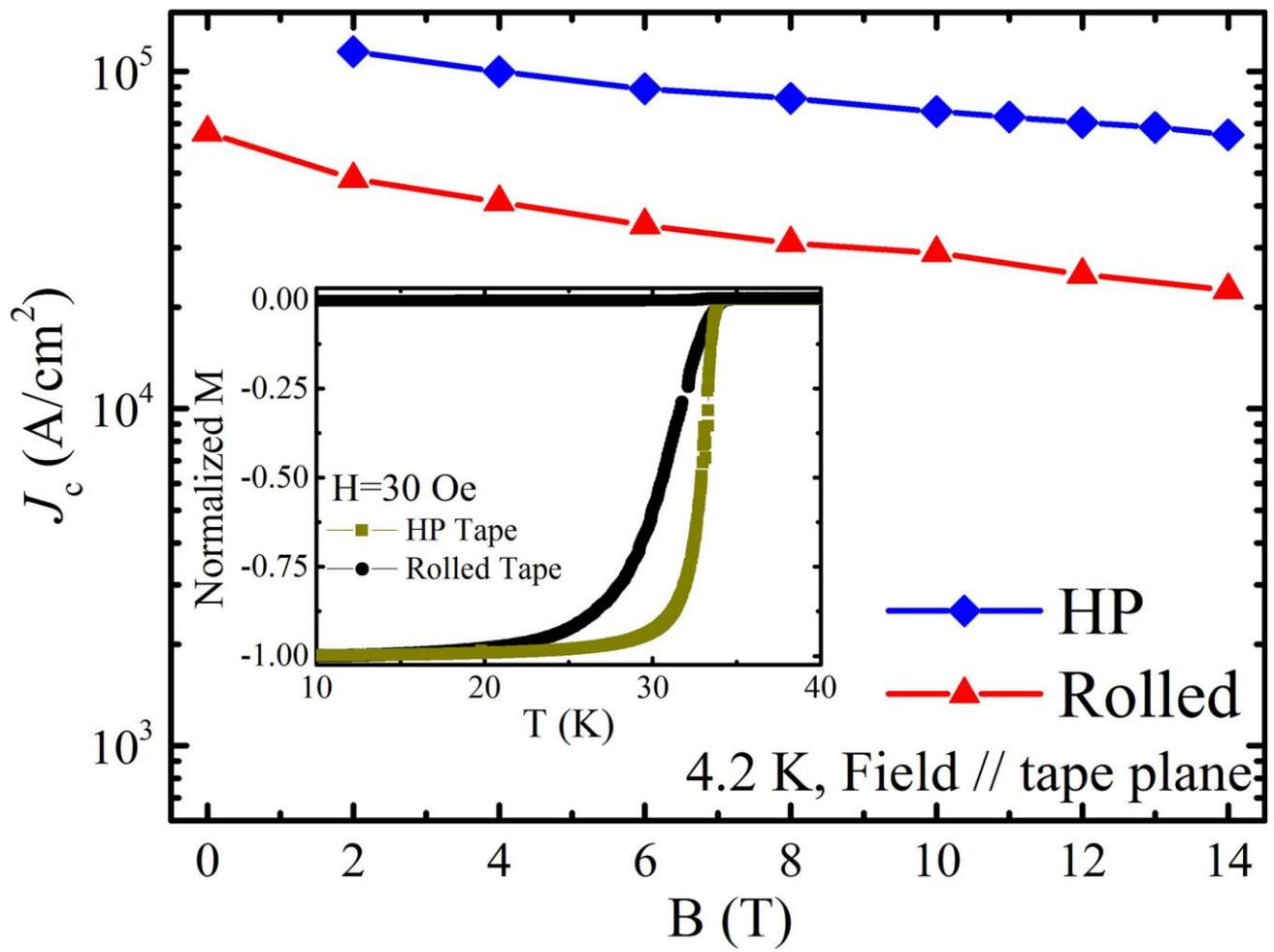

Figure 2

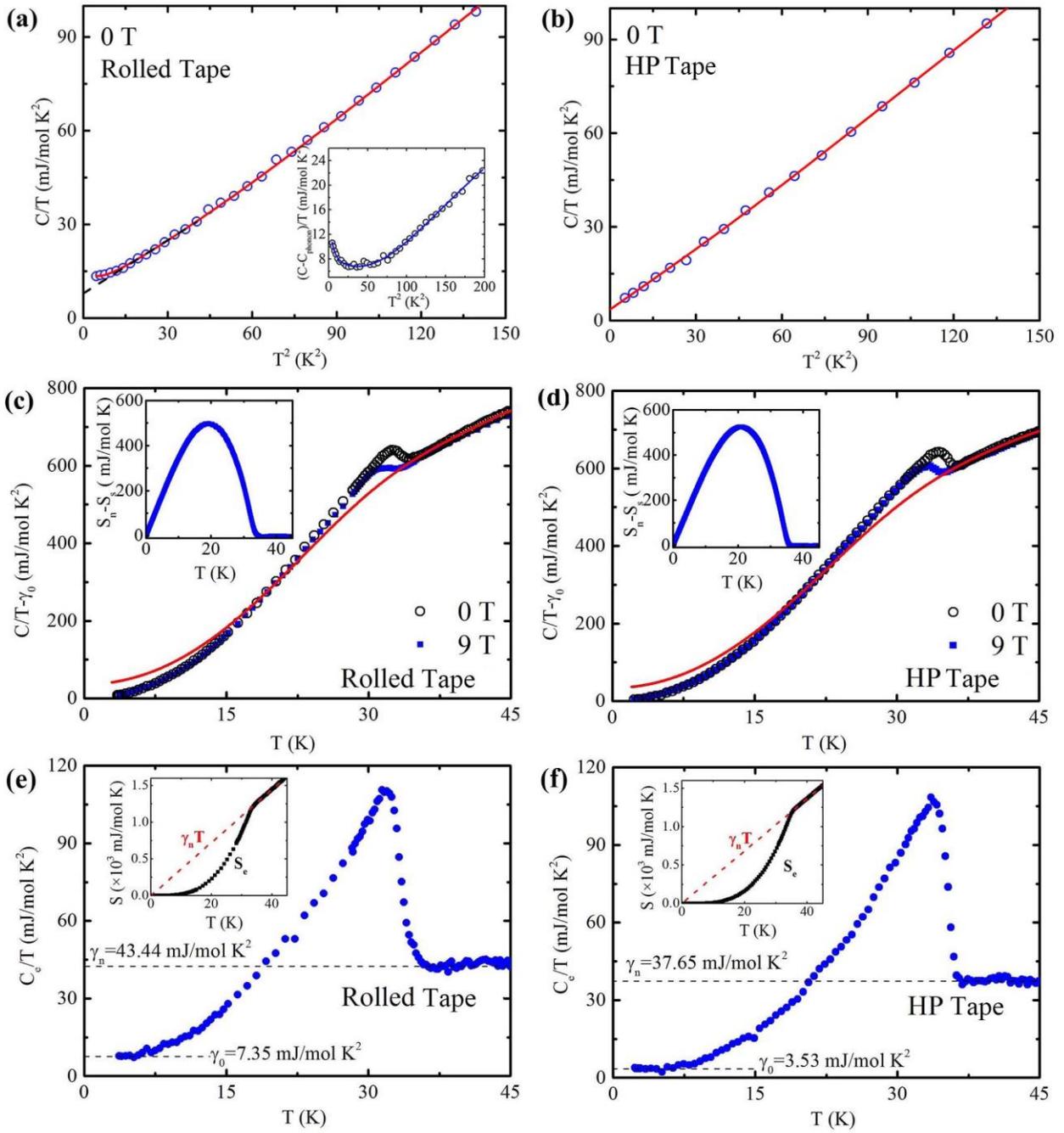

Figure 3

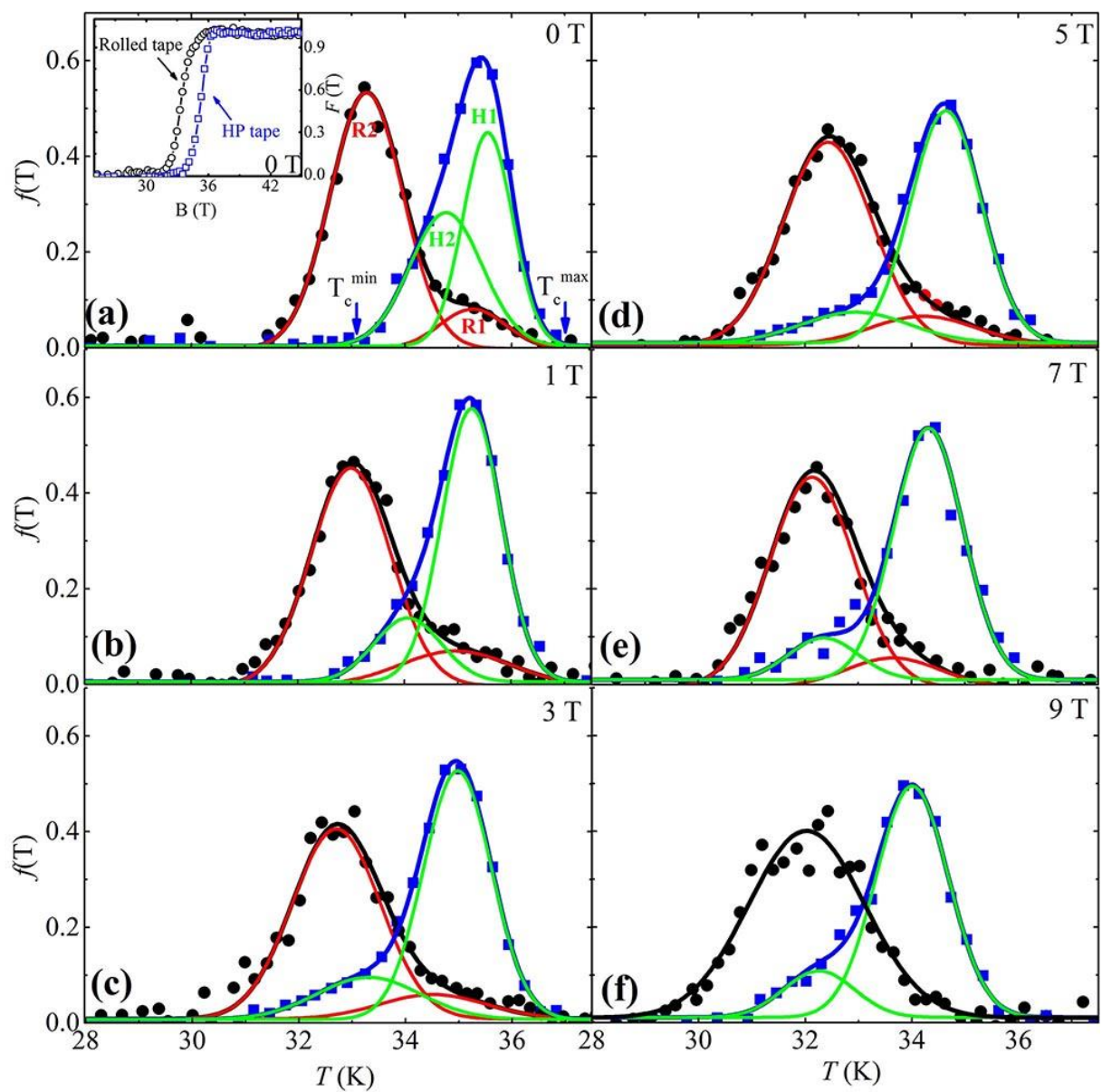

Figure 4

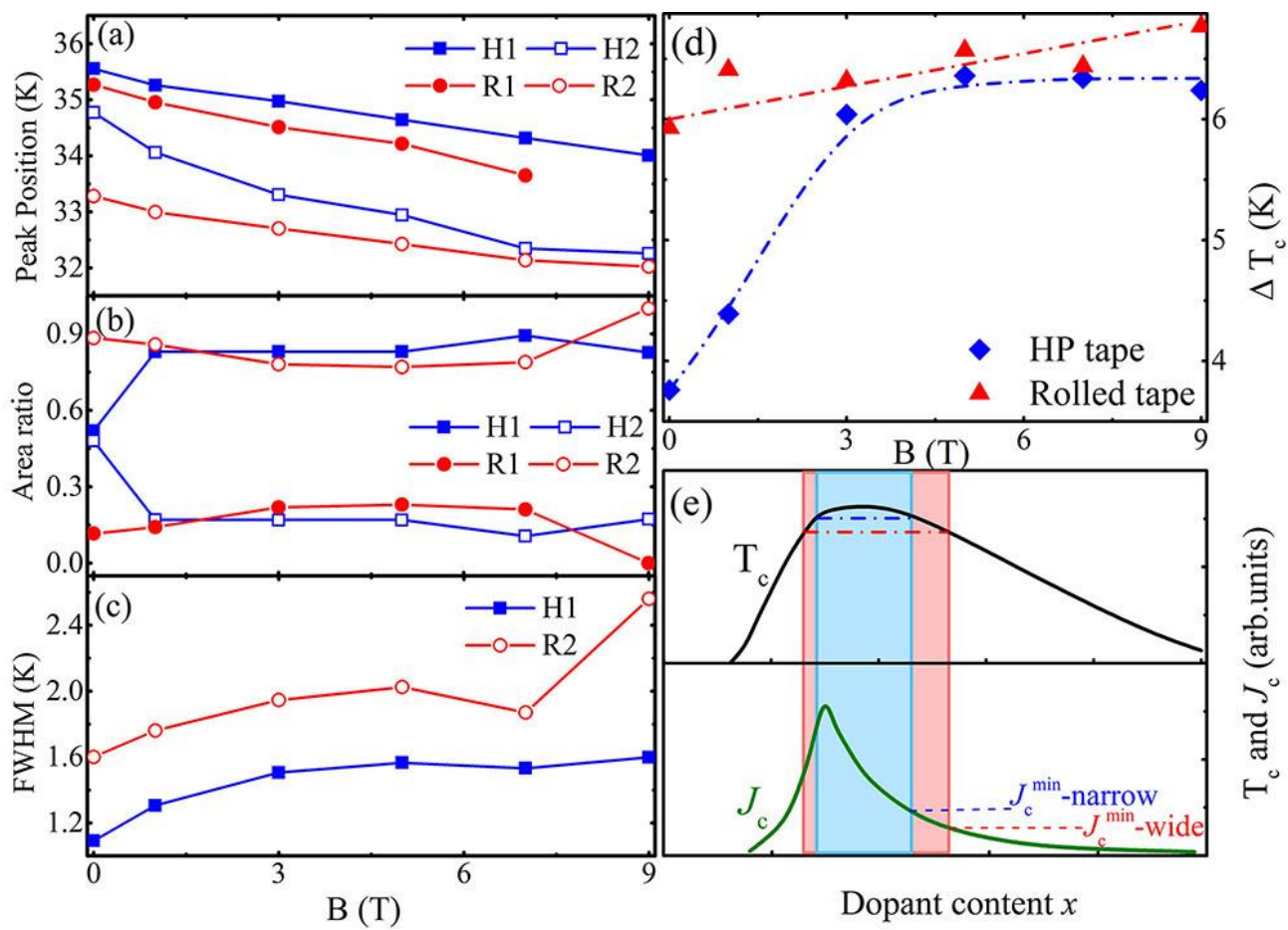